\documentclass[showpacs,prc,nofootinbib,twocolumn,floatfix]{revtex4}

\usepackage{graphicx}
\usepackage{latexsym}
\usepackage{hhline}

\sloppypar

\newcommand{\be}{\begin{equation}}
\newcommand{\ee}{\end{equation}}
\newcommand{\bea}{\begin{eqnarray}}
\newcommand{\eea}{\end{eqnarray}}

\newcommand{\mct}{{\mathcal T}}

\newcommand{\vt}{\vartheta}
\newcommand{\vp}{\varphi}

\newcommand{\foh}{\frac{1}{2}}
\newcommand{\fth}{\frac{3}{2}}

\newcommand{\ra}{\rightarrow}

\renewcommand{\vec}[1]{\mbox{\boldmath $#1 \!\!$ \unboldmath}}

\newcommand{\ibid}[3]{\textit{ibid.} {\bf #1}, #2 (#3)}
\newcommand{\JPR}[3]{Phys. Rev. {\bf #1}, #2 (#3)}
\newcommand{\JPRC}[3]{Phys. Rev. C {\bf #1}, #2 (#3)}
\newcommand{\JPRD}[3]{Phys. Rev. D {\bf #1}, #2 (#3)}
\newcommand{\JNP}[3]{Nucl. Phys. {\bf #1}, #2 (#3)}
\newcommand{\JAP}[3]{Ann. Phys. (N.Y.) {\bf #1}, #2 (#3)}
\newcommand{\JEPJA}[3]{Eur. Phys. J. A {\bf #1}, #2 (#3)}
\newcommand{\JEPJC}[3]{Eur. Phys. J. C {\bf #1}, #2 (#3)}
\newcommand{\JNCA}[3]{Nuovo Cimento A {\bf #1}, #2 (#3)}

\begin{document}

\title{$\pi N \ra \omega N$ in a coupled-channel approach} 

\author{G. Penner}
\email{gregor.penner@theo.physik.uni-giessen.de}
\author{U. Mosel}
\affiliation{Institut f\"ur Theoretische Physik, Universit\"at Giessen, D-35392
Giessen, Germany}

%\date{\today}

\begin{abstract}
We describe the $\pi N \ra \omega N$ cross section from threshold 
to a center of mass energy of $2$ GeV in a unitary coupled-channel
model and analyze it in terms of rescattering and resonance
excitations. The amplitude is mainly composed of $D_{13}$, $P_{13}$,
and $P_{11}$ contributions, where the $D_{13}$ dominates over the
complete considered energy range. We also outline the generalization of 
the standard partial-wave formalism necessary for the decomposition of
the $\omega N$ final state. 
\end{abstract}


\maketitle

\section{Introduction}

The reliable extraction of nucleon resonance properties from experiments 
where the nucleon is excited via either hadronic or electromagnetic 
probes is one of the major issues of hadron physics. The goal is 
to be finally able to compare the extracted masses and partial decay widths 
to predictions from lattice QCD (e.g., \cite{flee}) and/or quark models 
(e.g., \cite{capstick,riska}).

With this aim in mind we developed in \cite{feusti} a unitary
coupled-channel effective Lagrangian model that already incorporated
the final states $\gamma N$, $\pi N$, $2\pi N$, $\eta N$, and $K
\Lambda$ and was used for a simultaneous analysis of all avaible
experimental data on photon- and pion-induced reactions on the
nucleon. 

\begin{figure}
  \begin{center}
    \parbox{86mm}{\includegraphics[width=86mm]{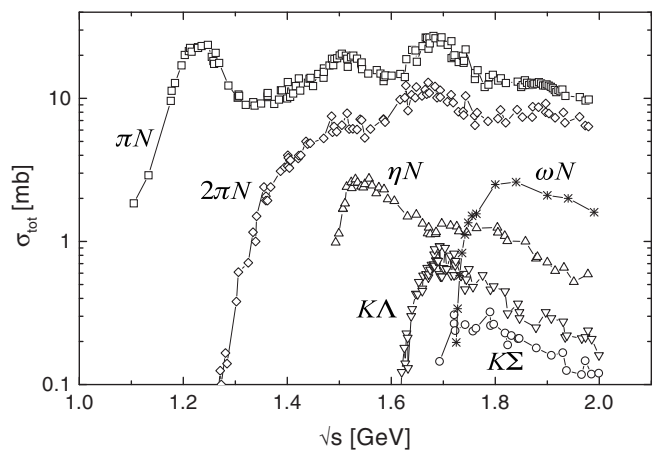}}
    \caption{Total cross sections for the reactions
      $\pi^-p \rightarrow X$ with $X$ as given in the figure. All data
      are from ref. \cite{landolt}; the lines are to guide the
      eye. \label{alltotals}} 
  \end{center}
\end{figure}

\begin{figure}
  \begin{center}
    \parbox{86mm}{\includegraphics[width=86mm]{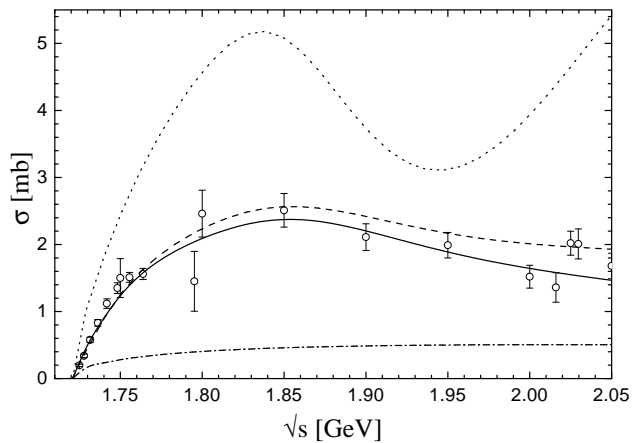}}
    \caption{$\pi^- p \ra \omega n$ total cross
      section. Solid line: full calculation. Dashed line: calculation
      with $g_{NN \omega} = 7.98$, $\kappa_{NN \omega} =
      -0.12$. Dotted line: no rescattering. Dash-dotted line: nucleon
      contribution ignoring rescattering. For the data references, see
      Sec. \ref{expdata}.\label{pototrescatt}} 
  \end{center}
\end{figure}

In an extension of the model to higher c.m. 
energies, i.e., up to center-of-mass energies of $\sqrt s = 2$ GeV for
the investigation of higher and so-called hidden nucleon resonances,
the consideration of other final states becomes unavoidable and hence the
model is extended to also include $\omega N$ and $K \Sigma$. As can be
seen from Fig. \ref{alltotals} 
for $\sqrt s > 1.7$ GeV it is mandatory to take into account the 
$\omega N$ state in a unitary model. Furthermore, $\omega$ production
on the nucleon represents 
a possibility to project out $I = \foh$ resonances in the reaction 
mechanism. However, the $\omega N$ channel resisted up to now a theoretical
description in line with experiment. Especially the inclusion of
nucleon Born contributions \cite{klingl} overestimated the data at
energies above $1.77$ GeV and only either the neglect of these diagrams 
\cite{post,friman} or very soft form factors \cite{titov} led to a
rough description of the experimental data\footnote{Note 
  that Ref. \cite{titov} did not use the correct experimental data,
  but followed the claim of Ref. \cite{hanhart99}; see Sec.
  \ref{expdata}.}. However, 
none of these models included rescattering effects or a detailed
partial-wave analysis of interference effects. As recently pointed out 
\cite{nstar2001} both lead to strong modifications of the observed
cross section; see also Fig. \ref{pototrescatt}. 

The aim of this paper is to present the results of $\pi N \ra \omega
N$ within a coupled-channel model that simultaneously describes all
pion induced data for $\pi N$, $2\pi N$, $\eta N$, $K \Lambda$, $K
\Sigma$, and $\omega N$. Hence this analysis differs from all  
other investigations of $\pi N \ra \omega N$ in two respects: First, a
larger energy region is considered, which also means there are more
restrictions from experiment, and second, the reaction process is
influenced by all other channels and vice versa. This leads to strong
constraints in the choice of $\omega N$ contributions and it is
therefore possible to extract them more reliably.  

We start with a short review of the model of
ref. \cite{feusti} in Sec. \ref{model}, where we also present the
way the $\omega N$ final state is included. As a result of the
$\omega$ intrinsic spin the inclusion of this final state requires an 
extension of the standard partial-wave decomposition (PWD) method
developed for $\pi N/\gamma N \rightarrow \pi N$ and $\gamma N
\rightarrow \gamma N$ (see, e.g., \cite{feusti}). Such an extension is
provided in Sec. \ref{omega}. In Sec. \ref{expdata} our
calculations are compared to the available experimental data and we
conclude with a summary.

\section{\label{model}Model}

The scattering equation that needs to be solved is the Bethe-Salpeter
(BS) equation for the scattering amplitude:
\bea
M(p',p;\sqrt s) &=& V(p',p;\sqrt s) 
\nonumber \\
&& + \int \frac{\mathrm d^4 q}{(2\pi)^4} V(p',q;\sqrt s) \nonumber \\ 
&& \; \; \; \; \times G_{BS}(q;\sqrt s) M(q,p;\sqrt s) \; .
\eea
Here, $p$ ($k$) and $p'$ ($k'$) are the incoming and outgoing baryon
(meson) four-momenta. After splitting up the two-particle BS propagator 
$G_{BS}$ into its real and imaginary parts, one can introduce the $K$-matrix 
via (in a schematical notation) $K = V + \int V \mathrm{Re} G_{BS}
M$. Then $M$ is given by $M = K + i \int M \mathrm{Im} G_{BS}
K$. Since the imaginary part of $G_{BS}$ just contains its on-shell
part, the reaction matrix $T$, defined via the scattering matrix 
$S = 1 + 2 i T$, can now be calculated from $K$ after 
a PWD in $J$, $P$, and $I$ via matrix inversion:
\be
T (p',p;\sqrt s) = \frac{K(p',p;\sqrt s)}{1 - i K(p',p;\sqrt s)} .
\ee
Hence unitarity is fulfilled as long as $K$ is Hermitian. For 
simplicity we apply the so-called $K$-matrix Born approximation, which 
means that we neglect the real part of $G_{BS}$ and thus $K$ reduces 
to $K = V$. The validity of this approximation was tested by Pearce
and Jennings \cite{pearce}. By fitting the $\pi N$ elastic phase
shifts also using other intermediate propagators for $G_{BS}$ these
authors found no significant differences in the extracted parameters.

The potential $V$ is built up by a sum of $s$-, $u$-, and $t$-channel 
Feynman diagrams by means of effective Lagrangians which can be found in 
\cite{feusti}. The background (nonresonant) contributions to the
amplitudes are not added ``by hand'', but are consistently
created by the $u$- and $t$-channel diagrams. Thus the number 
of parameters is greatly reduced. This holds true for the reaction
$\pi^- p \ra \omega N$ in the same way, where we also allowed for the
nucleon  Born diagrams and a $\rho$ exchange in the $t$ channel. In
our model the following 14 resonances are included: $P_{33}(1232)$,
$P_{11}(1440)$, 
$D_{13}(1520)$, $S_{11}(1535)$, $P_{33}(1600)$, $S_{31}(1620)$, 
$S_{11}(1650)$, $D_{33}(1700)$, $P_{11}(1710)$, $P_{13}(1720)$,
$P_{31}(1750)$, $P_{13}(1900)$, $P_{33}(1920)$, and a $D_{13}(1950)$
(as in \cite{feusti,bennhold}) which is listed as
$D_{13}(2080)$ by the Particle Data Group \cite{pdg}\footnote{Note 
  that the mass of this resonance as given by the references in
  \cite{pdg} ranges from $1.8$ to $2.08$ GeV.}. 

The resonance $\omega N$ Lagrangians have been chosen as a compromise
of an extension of the usual $RN\gamma$ transitions \cite{feusti} 
[for vector meson dominance (VMD) reasons] and the compatibility with
other $RN$ vector meson couplings used in the literature
\cite{riska,titov,postrho}; the latter point is discussed in
Sec. \ref{expdata}. For the spin-$\foh$ resonances we apply the same
$\omega N$ Lagrangian as for the nucleon ($\omega N \ra R$): 
\be
{\mathcal L} = -\bar R 
\left( \begin{array}{c} 1 \\ - i \gamma_5 \end{array} \right)
\left( g_1 \gamma_\mu - \frac{g_2}{2 m_N} \sigma_{\mu \nu}
  \partial^\nu_\omega \right) N \omega^\mu \; ,
\label{lagr12}
\ee
where the first coupling is the same one as in \cite{riska,titov}
since the $\omega$ is polarized such that $k'_\mu \omega^\mu = 0$. For
the spin-$\fth$ resonances we use
\bea
{\mathcal L} &=& -\bar R^\mu 
\left( \begin{array}{c} i \gamma_5 \\ 1 \end{array} \right)
\left( \frac{g_1}{2m_N} \gamma^\alpha + i \frac{g_2}{4 m_N^2} 
  \partial^\alpha_N + i \frac{g_3}{4 m_N^2} 
  \partial^\alpha_\omega \right) \nonumber \\
&& \times 
\left( \partial^\omega_\alpha g_{\mu \nu} - \partial^\omega_\mu
  g_{\alpha \nu} \right) N \omega^\nu \; .
\label{lagr32}
\eea
In both equations the upper operator ($1$ or $i \gamma_5$) corresponds
to a positive- and the lower one to a negative-parity resonance. For
positive-parity spin-$\fth$ resonances the first coupling is also
the same as used in \cite{riska,titov}; for negative
parity a combination of our first two couplings corresponds on shell
to theirs. The above couplings have also been applied in
\cite{postrho} in calculations of the $\rho$ spectral function.

Each vertex is multiplied with a cutoff function as in \cite{feusti}:
\be
F(q^2) = \frac{\Lambda_q^4}{\Lambda_q^4 + (q^2 - m_q^2)^2} \; ,
\ee
where $m_q$ ($q^2$) denotes the mass (four-momentum squared) of the
off-shell particle. To reduce the number of parameters the cutoff
value $\Lambda_q$ is chosen to be 
identical for all final states. We only distinguish between the
nucleon cutoff ($\Lambda_N$), the spin-$\foh$ ($\Lambda_\foh$) and
spin-$\fth$ ($\Lambda_\fth$) resonance cutoffs, and the $t$-channel
cutoff ($\Lambda_t$), i.e., only four different cutoff parameters.

From the couplings in Eqs. (\ref{lagr12}) and (\ref{lagr32}) the
helicity decay amplitudes of the resonances to $\omega N$ can be
deduced:
\bea
A^{\omega N}_{\foh} 
&=& \mp \frac{\sqrt{E_N \mp m_N}}{\sqrt{m_N}} \left( g_1 + g_2
  \frac{m_N \pm m_R}{2 m_N} \right) \; , \\
A^{\omega N}_0 
&=& \mp \frac{\sqrt{E_N \mp m_N}}{m_\omega \sqrt{2 m_N}} \left( 
  g_1 (m_N \pm m_R) + g_2 \frac{m_\omega^2}{2 m_N} \right) \; , 
\nonumber 
\eea
for spin-$\foh$ and 
\begin{widetext}
\bea
A^{\omega N}_{\fth} 
&=& - \frac{\sqrt{E_N \mp m_N}}{\sqrt{2 m_N}}\frac{1}{2m_N}
\left( g_3 \frac{m_\omega^2}{2 m_N} - g_1 (m_N \pm m_R) +
  g_2 \frac{m_R^2 - m_N^2 - m_\omega^2}{4 m_N}
  \right)  \; , \nonumber \\
A^{\omega N}_{\foh} 
&=& \pm \frac{\sqrt{E_N \mp m_N}}{\sqrt{6 m_N}}\frac{1}{2m_N}
\left( g_3 \frac{m_\omega^2}{2 m_N} \pm g_1 \frac{m_N (m_N
    \pm m_R) - m_\omega^2}{m_R} + g_2 \frac{m_R^2 - m_N^2 - m_\omega^2}{4 m_N}
  \right)  \; , \nonumber \\
A^{\omega N}_0 
&=& \pm m_\omega \frac{\sqrt{E_N \mp m_N}}{\sqrt{3 m_N}}\frac{1}{2m_N}
\left( g_1 \mp g_2 \frac{m_R^2 + m_N^2 - m_\omega^2}{4 m_R
    m_N} \mp g_3 
  \frac{m_R^2 - m_N^2 + m_\omega^2}{4 m_R m_N} \right) \; , 
\eea
\end{widetext}
for spin-$\fth$ resonances. Again, the upper sign holds for positive-
and the lower for negative-parity resonances. The lower indices
correspond to the resonance helicities and are determined by the
$\omega$ and nucleon spin $z$ components: $\fth$: $1 + \foh = \fth$,
$\foh$: $1 - \foh = \foh$, and $0$: $0 + \foh = \foh$. The resonance 
$\omega N$ decay widths are then given by 
\be
\Gamma^{\omega N} = \frac{2}{2J+1}\sum\limits_{\lambda = 0}^{\lambda = 
  +J} \Gamma^{\omega N}_\lambda \; , \hspace{4mm} 
\Gamma^{\omega N}_\lambda = \frac{\mathrm k' m_N}{2 \pi m_R} \left|
  A^{\omega N}_\lambda \right|^2 \label{heliwidth}
\ee
(upright letters denote the absolute value of the corresponding three-
momentum). As a result of the limited amount of experimental data (we
included 114 $\omega N$ data points in the fitting procedure;
cf. Sec. \ref{expdata}) we tried to minimize the set of parameters and
only varied a subset of the $\omega N$ coupling constants. This also
means that it is not possible to distinguish with certainty between
the different choices of the $RN\omega$ couplings, especially for
those resonances with only small contributions to $\omega N$. Only
more $\omega N$ data in the higher-energy region, i.e., above $\sqrt s
= 1.77$ GeV, and the inclusion of photoproduction data in the analysis 
\cite{moretocome} could shed more light on the situation. However, as
shown in Sec. \ref{expdata}, the choice of couplings presented in
the following allows a complete description of the angular and energy
dependences of the $\omega N$ production process. 

In the process of the fitting procedure we allowed for two different 
couplings ($g_1$ and $g_2$) to $\omega N$ for those resonances which
turned out to couple strongly to this final state, i.e.,
$P_{11}(1710)$, $P_{13}(1720)$, $P_{13}(1900)$, and $D_{13}(1950)$,
and one coupling ($g_1$) for the $S_{11}(1650)$. Since the usual
values for the $NN\omega$ couplings (cf. Ref. \cite{feusti} and
references therein) stem from different kinematical regimes than the
one examined here, we also allowed these two values to be varied
during the fitting procedure. But at the same time, the cutoff value
in the vertex form factor is not allowed to vary freely; instead, the
same value is used for all final states (see Sec. \ref{expdata}). It
is also important to notice that as a result of the coupled-channel
calculation, there are also constraints from all other 
channels that are compared to experimental data, leading to large
restrictions in the freedom of chosing the $\omega N$ contributions.

\section{\label{omega}$\omega$ Production}

Since the orbital angular momentum $\ell$ is not conserved in, e.g., 
$\pi N \rightarrow \omega N$, the standard PWD becomes inconvenient
for many of the channels that have to be included. Hence we use here a 
generalization of the standard PWD method which represents a tool to  
analyze any meson- and photon-baryon reaction on an equal, uniform footing. 

We start with the decomposition of the two-particle c.m. momentum
states ($\vec p = -\vec k$, $\mathrm p = |\vec p|$) into 
states with total angular momentum $J$ and $J_z = M$ \cite{jacobwick}:
\be
|\mathrm p JM,\lambda_k \lambda_p \rangle = 
  N_J \int e^{i(M-\lambda)\varphi} d^J_{M \lambda}(\vartheta) 
  |\mathrm p \vartheta \varphi,\lambda_k \lambda_p \rangle \mbox d \Omega,
\nonumber
\ee
where $\lambda_k$ ($\lambda_p$) is the meson (baryon) helicity and 
the $d^J_{M \lambda}(\vartheta)$ are Wigner functions. The
normalization $N_J$ is given by $\sqrt{(2J+1)/(4\pi)}$ and $\lambda =
\lambda_k - \lambda_p$. For the incoming c.m. 
state ($\vartheta_0 = \varphi_0 = 0$ $\Rightarrow$ $\ell = 0$) one gets 
$\langle JM,\lambda_k \lambda_p | \vartheta_0 \varphi_0 ,\lambda_k 
\lambda_p \rangle \sim \delta_{M \lambda}$, and one can 
drop the index $M$. By using the parity property \cite{jacobwick} 
$\hat P |J,\lambda \rangle = \eta_k \eta_p 
(-1)^{J-s_k-s_p} |J,-\lambda \rangle$, where $\eta_k$ and $\eta_p$
($s_k$ and $s_p$) are the intrinsic parities (spins) of the two particles, 
the construction of normalized states with parity $(-1)^{J \pm
  \foh}$ is straightforward:
\begin{eqnarray}
|J,\lambda;\pm \rangle &\equiv& 
\frac{1}{\sqrt 2} 
\left( |J,+\lambda\rangle \pm \eta |J,-\lambda \rangle \right)
\nonumber \\
\Rightarrow
\hat P |J,\lambda; \pm\rangle &=&  
(-1)^{J\pm \foh} |J,\lambda; \pm\rangle,
\end{eqnarray}
where we have defined
\be
\eta \equiv \eta_k \eta_p (-1)^{s_k+s_p+\foh} \; . \nonumber
\ee
They can be used to project out helicity amplitudes with parity
$(-1)^{J \pm \foh}$: 
\be
\mct^{J\pm}_{\lambda' \lambda} \equiv 
\langle J,\lambda';\pm | T | J,\lambda;\pm \rangle 
= \mct^J_{\lambda' \lambda} \pm \eta \mct^J_{\lambda' -\lambda}
\label{pariampli} \; ,
\ee
with
\bea
\mct^J_{\lambda'\lambda} (\sqrt s) &\equiv& 
\langle \lambda' | T^J(\sqrt s) | \lambda \rangle \nonumber \\
&=&
2 \pi \int \mathrm d (\cos \vt) d^J_{\lambda \lambda'} (\vt)
\langle \vt, \vp=0,\lambda' | T |0 0,\lambda \rangle \; .
\nonumber 
\eea
In eqn. (\ref{pariampli}) we have used, that for parity conserving
interactions $T = \hat P^{-1} T \hat P$:
\be
\langle J, -\lambda' | T | J, -\lambda \rangle 
= \eta (\eta')^{-1} \langle J, \lambda' | T | J, \lambda \rangle \; .
\ee
The helicity amplitudes $\mct^{J\pm}_{\lambda' \lambda}$ have
definite, identical $J$ and definite, but opposite $P$. As is quite
obvious this method is valid for any meson-baryon final state
combination, even such cases as, e.g., $\omega N \rightarrow \pi
\Delta$. In the case of $\pi N \rightarrow \pi N$ the
$\mct^{J\pm}_{\lambda' \lambda}$ coincide with the conventional
partial-wave amplitudes: $\mct^{J\pm}_{\foh \foh} \equiv
\mct_{\ell\pm}$. 

\section{\label{expdata}Comparison with Experiment}

For the fitting procedure we modified the data set used in
Ref. \cite{feusti} in the following way.

For $\pi N \ra \pi N$ we used the updated single-energy partial-wave
analysis SM00 \cite{SM00}. For $2 \pi N$, $\eta N$, and $K
\Lambda$ we continue to use the same database as in \cite{feusti};
however, for $\eta N$ the data from \cite{morrison} and for $K
\Lambda$ the data from \cite{lambdaexp} were added. For $K \Sigma$
production we used the total cross section, angle-differential cross
section, and polarization data from \cite{haba} and from the references
to be found in \cite{landolt}.

Furthermore, we have included all the $\pi N \ra \omega N$
data in the literature \cite{binnie,keyne,karami,danburg}. 
At this point we wish to stress that we do not follow the authors
of Refs. \cite{hanhart99,hanhart01} to ``correct'' the Karami
\cite{karami} data. The authors of \cite{hanhart99} have claimed that
the method used in \cite{binnie,keyne,karami} to extract the two-body
cross section from the count rates was incorrect. However, a careful
reading of Ref. \cite{binnie} reveals that the two-body cross sections
were indeed correctly deduced and the peak region of the $\omega$
spectral function is well covered even at energies close to the
$\omega$ production threshold. The conclusion of Ref. \cite{hanhart99}
can be traced back to the incorrect reduction of the integration over
the $\omega$ spectral function to the experimental averaging over the
outgoing neutron c.m. momentum interval binning; a detailed discussion 
can be found in \cite{fightforkarami}. See also the discussion about
the $\pi N$ inelasticities below.

The results presented in the following are from ongoing calculations
to describe the data of all channels simultaneously (cf. Table
\ref{chisquares}). 
\begin{table}
  \begin{center}
    \begin{tabular}
      {ccccccc}
      \hhline{=======}
      Total & $\pi N$ & $2\pi N$ & $\eta N$ & $K \Lambda$ & $K \Sigma$ &
      $\omega N$ \\
      \hline
      3.08 & 3.78 & 6.95 & 1.78 & 2.05 & 2.43 &  2.53 \\
      \hhline{=======}
    \end{tabular}
  \end{center}
  \caption{$\chi^2$ per degree of freedom from the present calculation 
    for $\pi N \ra X$ with $X$ as given in the table. 
    \label{chisquares}}
\end{table}
The coupling set used for the presented results leads to an overall $\chi^2$
of 3.08 per degree of freedom (by comparison to a total of $2360$ data
points).  

\begin{figure}
  \begin{center}
    \parbox{86mm}{\includegraphics[width=86mm]{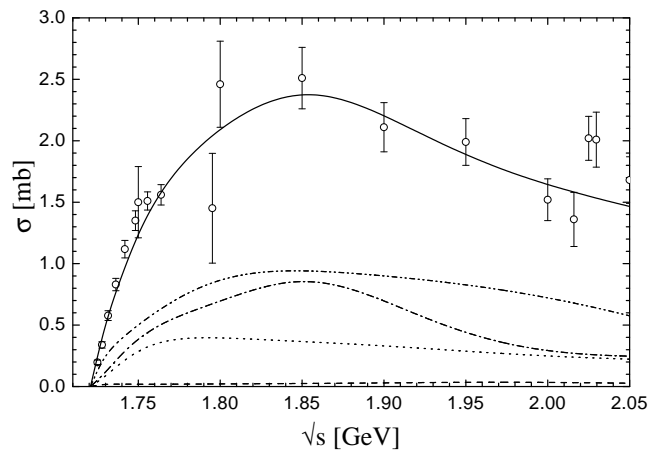}}
    \caption{$\pi^- p \ra \omega n$ total cross section. The
      contributions of various partial waves are given by
      $J^P=\foh^-(S_{11})$: dashed line; $\foh^+(P_{11})$: dotted line;
      $\fth^+(P_{13})$: dash-dotted line; $\fth^-(D_{13})$:
      dash-double-dotted line (in brackets the $\pi N$ notation is
      given). The sum of all partial waves is given by the solid 
      line. For the data references, see text.\label{totcrossom}}
  \end{center}
\end{figure}

\begin{figure*}
  \begin{center}
    \parbox{15cm}{\includegraphics[width=15cm]{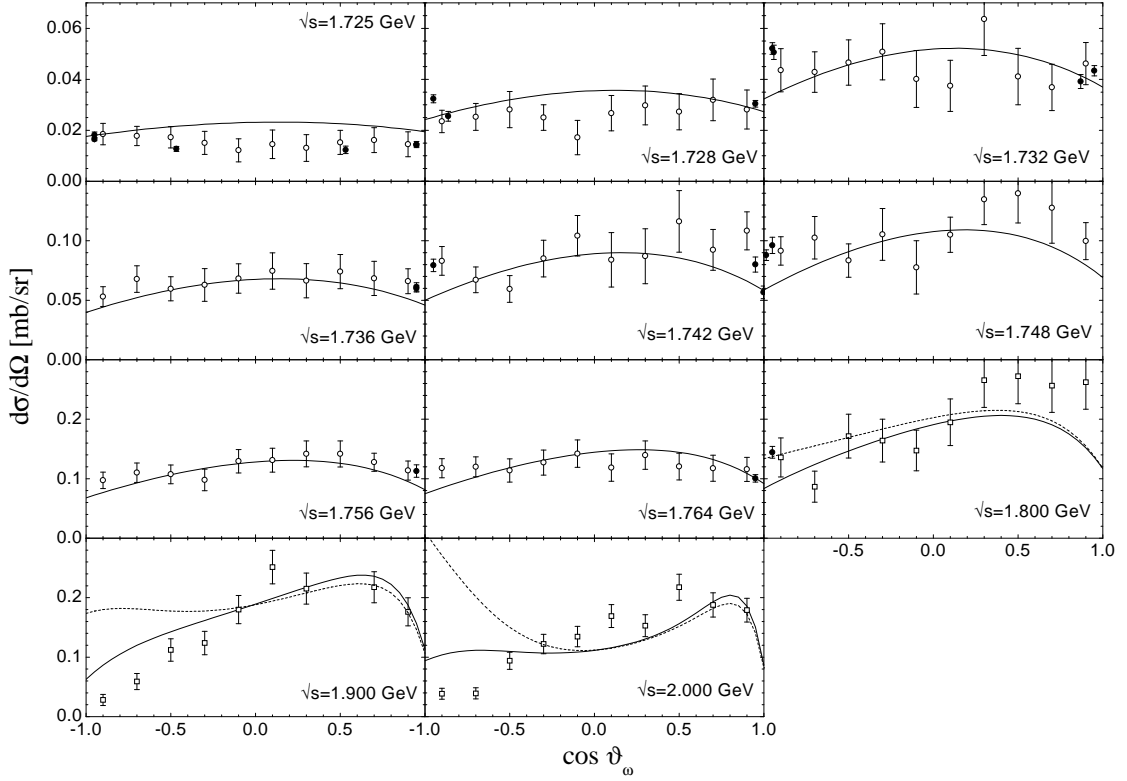}}
    \caption{$\pi^- p \ra \omega n$ differential cross section. Data
      are from $\bullet$: \cite{binnie,keyne}, $\circ$: \cite{karami}, 
      and $\Box$: \cite{danburg}. For the data points extracted from
      Ref. \cite{danburg} see text. At energies $\sqrt s \geq 1.8$ GeV
      also a calculation with $g_{NN \omega} = 7.98$, $\kappa_{NN
        \omega} = -0.12$ is shown (dashed line).\label{omegadif}} 
  \end{center}
\end{figure*}

As can be seen in Figs. \ref{totcrossom} and \ref{omegadif} 
our calculation is in line with all total and also with the 
differential $\omega N$ cross sections of
Refs. \cite{binnie,keyne,karami}\footnote{The total cross sections
  given in Refs. \cite{binnie,keyne} are actually angle-differential
  cross sections (mostly at forward and backward neutron c.m. angles)
  multiplied by $4\pi$.}. 
To get a handle on the angle-differential structure of the cross
section for energies $\sqrt s \geq 1.8$ GeV we also extracted 
angle-differential cross sections from the corrected cosine event 
distributions given in Ref. \cite{danburg} with the help of their
total cross sections. These data points strongly constrain
the nucleon $u$-channel contribution because of the decrease at
backward angles; see the end of this section. Moreover, for these
energies the contribution of the $\rho$ exchange contribution leads to
an increasing forward peaking behavior. 

The total $\omega N$ cross section (cf. Fig. \ref{totcrossom}) is
dominantly composed of two partial waves 
contributing with approximately the same magnitude
$J^P=\fth^-(D_{13})$ and $\fth^+(P_{13})$, and also a smaller
$\foh^+(P_{11})$ contribution, while the $\foh^-(S_{11})$ 
partial wave is almost negligible (in brackets the $\pi N$ notation is
given). The main contributions in these partial waves stem from the
$D_{13}(1950)$, the $P_{13}(1720)$, the nucleon, and the
$P_{11}(1710)$. The $D_{13}(1950)$ is especially interesting, since it
is only listed in the PDG \cite{pdg} at $2.08$ GeV, but was already
found as an important contribution in $\pi N$ and $K \Lambda$ channels
(cf. \cite{feusti,bennhold}) at around $1.95$ GeV. In our calculation 
it turns out to be an important production mechanism as well, in
particular at threshold. These findings are also contrary to the
conclusions drawn in \cite{karami}. Guided by their angle-differential 
cross sections they excluded any noticeable $J = \fth$ 
effects and deduced a production mechanism that is dominated by 
$J = \foh$ contributions. However, our coupled-channel
calculation shows that their angle-differential cross sections can
indeed be described by dominating $\fth^-$ and $\fth^+$
waves. Furthermore, since the data in all other channels (including
$\pi N$ inelasticities and $2\pi N$ partial wave cross sections in the 
isospin-$\foh$ partial waves; see below) are also very well described
in the $\omega N$ threshold region ($1.72$ GeV $<\sqrt s < 1.76$ GeV), 
our partial-wave decomposition of $\pi N \ra \omega N$ is on safe
grounds. 

\begin{figure}[t]
  \begin{center}
    \parbox{86mm}{\includegraphics[width=86mm]{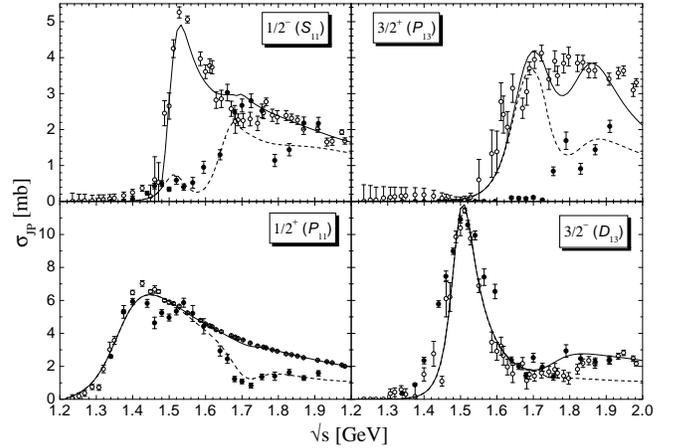}}
    \caption{$\pi N \ra \pi N$ inelastic ($\circ$ as
      extracted from SM00 \cite{SM00}, calculation: solid line) and
      $\pi N \ra 2 \pi N$ partial-wave cross sections ($\bullet$ as
      extracted by \cite{manley}, calculation: dashed line), both for
      $I=\foh$. For the discrepancy of $\pi N \ra 2 \pi N$ in
      the $\fth^+(P_{13})$ partial wave between $1.52$ and $1.725$
      GeV see text.\label{ppinp2i12}} 
  \end{center}
\end{figure}

\begin{figure}[t]
  \begin{center}
    \parbox{86mm}{\includegraphics[width=86mm]{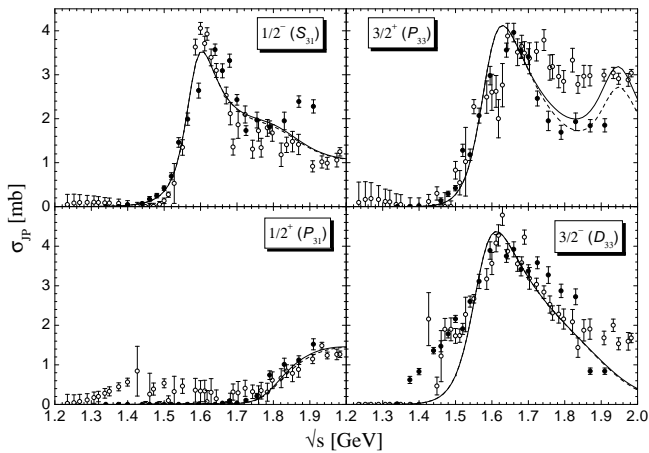}}
    \caption{$\pi N \ra \pi N$ inelastic ($\circ$ as
      extracted from SM00 \cite{SM00}, calculation: solid line) and
      $\pi N \ra 2 \pi N$ partial-wave cross sections ($\bullet$ as
      extracted by \cite{manley}, calculation: dashed line), both for
      $I=\fth$.\label{ppinp2i32}} 
  \end{center}
\end{figure}

As a result of the coupled-channel calculation, the opening of the
$\omega N$ channel also becomes visible in the inelasticity of the
$\pi N \ra \pi N$ channel. In figs. \ref{ppinp2i12} and
\ref{ppinp2i32}
the $\pi N \ra \pi N$ inelastic
\be
\sigma^{in}_{IJ\pm} = \frac{4\pi}{\mathrm k^2} \left( J+\foh \right)
\left( \mathrm{Im} \mct^{IJ\pm}_{\foh \foh} - \left|
    \mct^{IJ\pm}_{\foh \foh} \right|^2 \right)
\ee
and the $\pi N \ra 2 \pi N$ partial-wave cross sections
\be
\sigma^{2\pi}_{IJ\pm} = \frac{4\pi}{\mathrm k^2} \left( J+\foh \right)
\left| \mct^{IJ\pm}_{\foh \foh} \right|^2
\ee
are plotted together with experimental data from SM00 \cite{SM00} and
\cite{manley}. An $IJ^P=\foh \foh^-$ or $\foh \fth^-$ wave contribution 
in the order of $\sigma_{\omega N} \geq 3$ mb for $1.72$ GeV $\leq
\sqrt s \leq$ $1.74$ GeV as claimed in \cite{hanhart99,hanhart01}
would also be in contradiction with inelasticities extracted from 
$\pi N \ra \pi N$ partial waves: The $\foh \foh^-$ inelasticity around
the $\omega N$ threshold is already saturated by the $2 \pi N$ and $K
\Sigma$ channels; a large $\omega N$ contribution would spoil the
agreement between calculation and experiment; the $\foh \fth^-$
inelasticity allows only $\sigma^{\omega N}_{\foh \fth^-} \leq
\sigma^{in}_{\foh \fth^-} - \sigma^{2\pi}_{\foh \fth^-} \approx 1$ mb
in this energy region.

At this point a remark on the $IJ^P=\foh \fth^+$ inelasticity between
$1.52$ and $1.725$ GeV is in order. This inelasticity grows up to $4$
mb below the $\omega N$ threshold, while the $2\pi N$ partial-wave
cross section extracted by \cite{manley} is still zero. At the same
time all total cross sections from other open inelastic channels
($\eta N$, $K \Lambda$, and $K \Sigma$) add up to significantly less
than $4$ mb. This indicates that either the extracted $2 \pi N$
partial wave cross section is not correct in the $\foh \fth^+$ partial 
wave or another inelastic channel (i.e., a $3 \pi N$ channel)
contributes significantly to this partial wave\footnote{The same
  problem was observed in a resonance parametrization of $\pi N \ra
  \pi N$ and $\pi N \ra 2 \pi N$ \cite{manley92}.}. Note that we only 
observe this effect in this partial wave and are also able to describe
the inelasticity and the $2 \pi N$ data above the $\omega N$ threshold 
in the $\foh \fth^+$ partial wave. Therefore, we did not introduce an
additional final state but effectively neglected the $\foh \fth^+$ $2
\pi N$ data points in the energy region between $1.52$ and $1.725$
GeV.

\begin{figure}[t]
  \begin{center}
    \parbox{86mm}{\includegraphics[width=86mm]{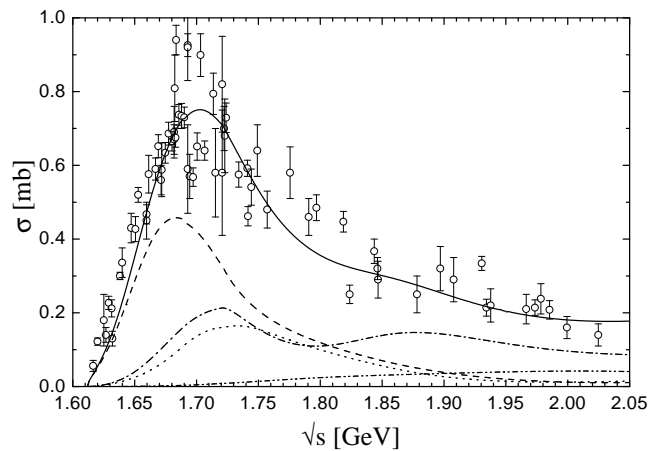}}
    \caption{$\pi^- p \ra K^0 \Lambda$ total cross section. See
      Fig. \ref{totcrossom} for the notation. For the data references,
      see text.\label{totcrosslam}}
  \end{center}
\end{figure}

\begin{figure}[t]
  \begin{center}
    \parbox{76mm}{\includegraphics[width=76mm]{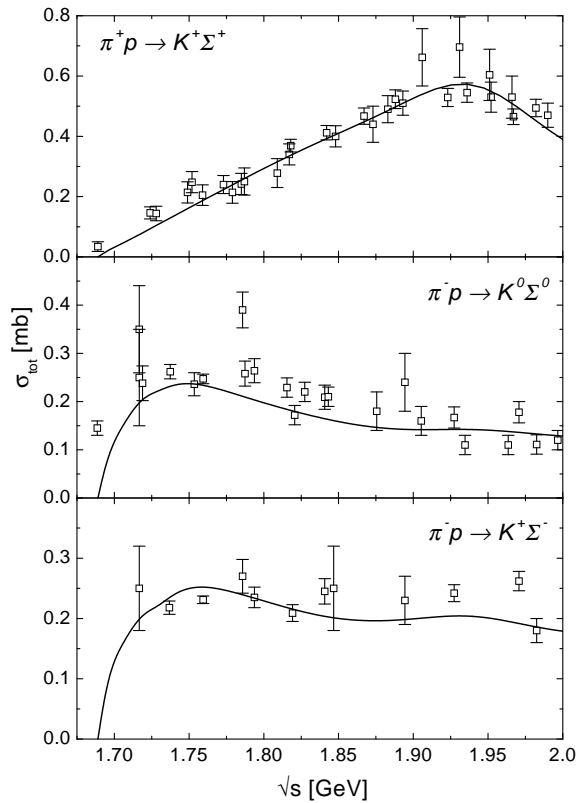}}
    \caption{$\pi N \ra K \Sigma$ total cross sections.\label{pstot}}
  \end{center}
\end{figure}

Another coupled-channel effect shows up in the total 
$\pi^- p \ra K^0 \Lambda$ cross  section. As can be seen in
Fig. \ref{totcrosslam} 
this channel exhibits a resonancelike behavior
for energies $1.67$ GeV$\leq \sqrt s \leq 1.73$ GeV. However, this
structure is also caused by the opening of two new channels, which take
away the flux in the $\foh^-$ and $\fth^+$ partial waves. First,
around $1.69$ GeV the $K \Sigma$ channel opens up with a strong
$IJ^P=\foh \foh^-$ contribution. Second, around $1.72$ GeV $\omega N$
opens up with a small $\foh \foh^-$ but a strong $\foh \fth^+$
wave. The $\pi N \ra K \Sigma$ cross sections are shown in
Fig. \ref{pstot}. 
The pure $I=\fth$ channel $\pi^+ p \ra K^+ \Sigma^+$ is strongly dominated
by a $\fth \fth^+$ wave and also becomes visible in the $\fth \fth^+$
$\pi N$ inelasticity (cf. Fig. \ref{ppinp2i32}), while the other two
channels $\pi^- p \ra K^0 \Sigma^0/K^+ \Sigma^-$ show the strong
$\foh \foh^-$ wave rise just above threshold.

As mentioned above we also allowed for the nucleon 
Born contributions in $\pi N \ra \omega N$ usually leading to an
overestimation of the total cross section at higher energies. As can be 
seen in Fig. \ref{pototrescatt}, the inclusion of rescattering is
mandatory to be able to describe the energy-dependent behavior of the
total $\pi N \ra \omega N$ cross section: When we apply our best 
parameter set to a tree level calculation --- i.e., ``rescattering'' is
only taken into account via an imaginary part in the denominator of
the resonance propagators --- the calculation results in the dotted
line, which is far off the experimental data. This shows the
importance of ``off-diagonal'' rescattering such as $\pi N \ra \omega
N \ra \pi N$ or $\pi N \ra K \Lambda \ra \omega N$. 

The values of the $NN\omega$ couplings are mainly determined by the
backward angle-differential cross section at higher energies. During
the fitting procedure these couplings resulted in $g_1=4.50$ and
$\kappa=g_2/g_1=-0.70$. The total cross section exhibits almost the
same behavior when we use the values from \cite{feusti} ($g_1=7.98$ and
$\kappa=-0.12$; see the dashed line in Fig. \ref{pototrescatt}); however,
for energies above $\sqrt s=1.8$ GeV the angular dependence (see the
dashed line in Fig. \ref{omegadif}) is not in line with experiment
anymore. The $NN$-meson cutoff value used for all $s$- and
$u$-channel diagram vertices (hence also for the $NN\omega$ vertex)
resulted in $\Lambda_N = 1.15$ GeV.

For the other background contribution in the $\omega N$ production,
i.e., the $\rho$ exchange, we used the couplings $g_{\omega \rho \pi} =
2.056$ (extracted from the $\omega \ra \rho \pi \ra \pi^+ \pi^- \pi^0$
width), $g_{NN\rho}=5.56$, and $\kappa_{NN\rho}=1.58$ --- the
latter values were extracted from the fit and are the same as in
calculating $\pi N$ elastic scattering.

In Table \ref{omegawidth}
\begin{table}
  \begin{center}
    \begin{tabular}
      {llrrrrrr}
      \hhline{========}
      $L^{\pi N}_{2I2J}$ & M & $\Gamma_{tot}$ & $\Gamma^{\omega N}_0$
      & $\Gamma^{\omega N}_{\foh}$ & $\Gamma^{\omega
        N}_{\fth}$ & $\Gamma^{\omega N}$ & $\Gamma^{\omega
        N}$ of {\cite{capstick}} \\ 
      \hline 
      $S_{11}(1650)$ & 1677.5 & 177 & -0.224$^a$ & 0.0$^{ab}$ & -- & --
      & -- \\ 
      $P_{11}(1710)$ & 1786.3 & 686 & 76 & 69 & -- & 145 & $0.0 \;
      \scriptstyle {{+5.3}\atop{-0.0}}$ \\ 
      $P_{13}(1720)$ & 1722.5 & 252 & 0.05 & 0.11 & 1.18 & 0.67 & $0.0 \;
      \scriptstyle {{+1.7}\atop{-0.0}}$ \\ 
      $P_{13}(1900)$ & 1951.0 & 585 & 21 & 0 & 226 & 123 & $20.3 
      \scriptstyle {{+34.8}\atop{-0.0}}$ \\ 
      $D_{13}(1950)$ & 1946.0 & 948 & 162 &  0 & 289 & 226 & $39.7 
      \scriptstyle {{+56.3}\atop{-21.2}}$ \\ 
      \hhline{========}
    \end{tabular}
  \end{center}
  \caption{Masses, total, and $\omega N$ widths (see
    Eqn. (\ref{heliwidth})) for $I=\foh$ resonances coupling to $\omega
    N$. All values are  given in MeV. For the $\omega N$ widths of
    ref. \cite{capstick}, we also cite the upper and lower values of
    their extracted ranges. $^a$The couplings $g_1$, $g_2$ are
    given. $^b$Not varied in the fit; see text.\label{omegawidth}} 
\end{table}
the resonance properties of those resonances which couple to $\omega
N$ are presented. In contrast to \cite{capstick,riska,titov} we also
find strong contributions from the $P_{11}(1710)$ and the
$P_{13}(1720)$ resonances, where the latter one is located just above
the $\omega N$ threshold of $1.721$ GeV. Our extracted $P_{11}(1710)$
width is significantly larger than the PDG \cite{pdg} value of
$\approx 100$ MeV\footnote{Note that the width of this resonance as
  given by the references in \cite{pdg} ranges from $90$ to $480$
  MeV.}, but consistent with the value of $480 \pm 230$ MeV extracted
by a resonance parametrization of $\pi N \ra \pi N$ and $\pi N \ra 2
\pi N$ \cite{manley92}. The reason for these large differences is the
lack of a prominent resonant behavior in the upper energy region of 
the $P_{11}$ $\pi N \ra \pi N$ partial wave. Thus the extraction of
resonance parameters is not well constrained by $\pi N \ra \pi N$
alone. In our analysis the large width comes to about one-fourth from
$\omega N$ and the remainder is due to $2 \pi N$ ($268$ MeV), $\eta
N$ ($160$ MeV), and $K \Sigma$ ($71$ MeV). In the latter two channels
strong $P_{11}$ contributions are needed to describe the corresponding 
angle-differential cross sections and polarization observables.

We can also compare our 
$S_{11}(1650)$ and $P_{13}(1720)$ couplings to the one from
\cite{riska,titov} if we choose to take the same width for the
$P_{13}$, but only use the first coupling ($g_1$). While we find only
a small $S_{11}$ coupling of $g_1 = -0.22$, but a large value  of $g_1
= 29.3$ for the $P_{13}$, \cite{riska,titov} found $-2.56$ and $3.17$,
respectively. However, as is clear from the discussion above, a strong
$P_{13}$ and a small $S_{11}$ are mandatory results of  our
coupled-channel analysis. For the $P_{11}(1710)$, $P_{13}(1720)$, and
$D_{13}(1950)$ also a comparison to the VMD
predictions of Ref. \cite{post} is possible. The authors of
Ref. \cite{post} used different photon helicity amplitude analyses to
extract ranges for the $RN\omega$ transition couplings under the
assumption of strict VMD. Using their notation we find from our widths
the following couplings: $P_{11}(1710)$: $6.3$ ($0$--$1.22$),
$P_{13}(1720)$: $15.6$ ($0$--$5.0$), and $D_{13}(1950)$: $2.3$
($0$--$2.6$). In brackets, their VMD ranges are given. As a result of
the large uncertainties in the photon helicity amplitudes, which are
the input to the calculation of \cite{post}, it is impossible to draw
any conclusion on the validity of strict VMD for these resonances.

\section{Conclusions and Outlook}

In this paper we have included the $\omega N$ final state into our
coupled-channel model and have investigated whether it is possible to
find a way to describe the hadronic $\omega N$ data. The results of
our calculations show that for a description of the reaction $\pi^- p
\ra \omega N$ in line with experimental data a unitary,
coupled-channel calculation is mandatory, and the resulting amplitude
is mainly composed of $IJ^P = \foh \fth^-$ ($D_{13}$), $\foh \fth^+$
($P_{13}$), and $\foh \foh^+$ ($P_{11}$) contributions, where the
$\foh \fth^-$ dominates over the complete considered energy range.  

The next step in our investigation of nucleon resonance properties
within our coupled-channel $K$-matrix model will naturally be the
inclusion of photon induced data to further pin down the extracted
widths and masses. The results of this study and also more details
about the calculation presented here will be published soon
\cite{moretocome}. 

Furthermore, since the partial-wave formalism is now settled, the
inclusion of additional final states, in particular for a more 
sophisticated description of the $2\pi N$ final state, as $\rho N$ or 
$\pi \Delta$ is rather straightforward. Also, by the inclusion of
several, e.g., $\rho N$ final states with different masses $m_\rho$ the
width of the $\rho$ meson (and similarly for the $\Delta$) can also be
taken into account. Finally, investigations concerning the inclusion
of spin $\frac{5}{2}$ resonances are underway.

\begin{acknowledgments}
This work was supported by DFG and GSI Darmstadt.
\end{acknowledgments}

\end{document}